# Automatic Segmentation, Feature Extraction and Comparison of Healthy and Stroke Cerebral Vasculature

Aditi Deshpande[1], Nima Jamilpour[1], Bin Jiang[3], Chelsea Kidwell[2], Max Wintermark[3], Kaveh Laksari[1, 4]

[1]Department of Biomedical Engineering, University of Arizona
[2]Department of Neurology, University of Arizona
[3]Department of Radiology, Stanford University
[4]Department of Aerospace & Mechanical Engineering, University of Arizona

Abstract:

Accurate segmentation of cerebral vasculature and a quantitative assessment of cerebrovascular morphology is critical to various diagnostic and therapeutic purposes and is pertinent to studying brain health and disease. However, this is still a challenging task due to the complexity of the vascular imaging data. We propose an automated method for cerebral vascular segmentation without the need of any manual intervention as well as a method to skeletonize the binary volume to extract vascular geometric features which can characterize vessel structure. We combine a Hessian-based probabilistic vessel-enhancing filtering with an active-contour-based technique to segment magnetic resonance and computed tomography angiograms (MRA and CTA) and subsequently extract the vessel centerlines and diameters to calculate the geometrical properties of the vasculature. Our method was validated using a 3D phantom of the Circle-of-Willis region, demonstrating 84% mean Dice Similarity and 85% mean Pearson Correlation with minimal modified Hausdorff distance error (3 surface pixels at most). We applied this method to a dataset of healthy subjects and stroke patients and present a quantitative comparison between the extracted stroke and normal vasculature. We found significant differences in the geometric features including total length (2.88 ± 0.38 m for healthy and 2.20 ± 0.67 m for stroke), volume (40.18 ± 25.55 ml for healthy and 34.43 ± 21.83 ml for stroke), tortuosity (3.24 ± 0.88 rad/cm for healthy and 5.80 ± 0.92 rad/cm for stroke) and fractality (box dimension1.36 ± 0.28 for healthy vs. 1.69 ± 0.20 for stroke), among others, between the two groups ($p < 0.05$). The vascular network in stroke patients was found to have a higher tortuosity, fractality and lower values for total length, total volume and average branch length, as hypothesized. This technique can be applied on any imaging modality and can be used in the future to automatically obtain the 3D segmented vasculature for diagnosis and treatment planning of Stroke and other cerebrovascular diseases (CVD) in the clinic and also to study the morphological changes caused by various CVD.

## Introduction:

Cerebrovascular diseases (CVD) are a leading cause of death and disability in the US and worldwide with stroke being a major contributor (1,2). Assessing the structural changes in cerebral vasculature is pertinent to brain health and diagnosing and characterizing disease (3,4). Changes in vascular structure can indicate altered function and potential cerebral pathophysiology (5,6) and hence need to be characterized and quantified (4,7), which in turn requires deep knowledge of the normal healthy vascular geometry and morphology (8). It has been shown in the literature that altered vascular properties such as the vessel diameters, tortuosity and the branching pattern are closely correlated with cerebrovascular diseases such as atherosclerosis and stroke (9,10). After an ischemic stroke, the reduced blood flow causes a series of changes leading to structural remodeling of the vasculature (11). Studying these acute and chronic changes in vascular structure is key to understanding the underlying physiological mechanism of disease and cerebral function. A well segmented cerebral vasculature map is essential to visualize and quantify vessel occlusions, evaluate cerebral blood flow and perfusion, assess the extent of ischemia in stroke patients, and to detect and assess other cerebral vascular malformations such as aneurysms (2,12,13).

Furthermore, in neurosurgical planning, choosing the appropriate endovascular procedure and determining the best surgical plan necessitates a 3D segmented map of the vasculature (14–16).

For efficient diagnosis and treatment of stroke and other CVD, angiography imaging techniques such as Computed Tomography Angiography (CTA) and Magnetic Resonance Angiography (MRA) are routinely performed in the clinic as well as in acute hospital settings to visualize the blood vessels and flow of blood in the brain (2,17). These imaging techniques, however, only show 2D cross sectional slices which contain other anatomical structures as well as noise (18). This potentially leads to error in diagnosis due to the partial information regarding vascular structure without a 3D volume (2) . Despite the clinical need, there is still a lack of automated segmentation of patient-specific 3D cerebral vasculature and subsequent feature extraction due to the various challenges posed by this problem. Cerebral vessel geometries vary greatly in length, diameter and tortuosity, making vessel tracking and segmentation a complex multi-scale problem (19–21). Intensity inhomogeneities and inconsistent contrast exist due to flow velocity changes as well as varying imaging protocols (21,22). An overlap of bony anatomical structures in CT and white/grey matter in MR can distort the vascular imagery due to a shadow effect (19). Also, the smaller vessels whose diameters are at the scale of the highest achievable image resolution are extremely hard to detect (15). Due to all of these challenges, designing a robust segmentation method that works on all imaging modalities, remains the biggest challenge in visualization and assessment of 3D cerebral vasculature in the clinic (15,23). In the past few years, there has been a significant effort towards solving the problem of segmentation and an increasing number of approaches have been proposed in the literature (12,24). Most of these methods however have only been evaluated on a specific type of imaging data, such as either MRA or CTA (4,12,18,25). Thresholding-based methods that utilize either global or local thresholds tend to lose vessel pixels, resulting in inaccurate segmentations (24,26). Other methods need varying forms of manual interventions at different stages of segmentation or feature extraction (7,19). Also, many of the proposed methods in literature that use centerline tracking to detect the vessels require some form of manual 'seed points' or initialization due to inconsistencies in intensity along the vessels in different slices (12,15,24). Furthermore, geometric feature extraction for characterization of cerebral vessels has been scarcely reported in the literature (8) with only a few major attempts (4,7). However, these methods require manual intervention in certain stages during segmentation and feature extraction. Most other 3D visualization and analysis tools in literature utilize only global geometric features such as length and volume (27,28), even though local regional features provide more useful and targeted information on vasculopathies (29,30). Some of the major contributions to cerebral vessel segmentation in literature and their limitations are listed in Table 1.

To the best of our knowledge, there aren't any comprehensive studies which quantitatively compare the differences in cerebral vascular structure and geometry in major vascular pathological conditions such as stroke with healthy subjects, which is an essential investigation for understanding brain health and disease (4,18). Some studies have extracted vascular features of patients with intracranial arterial stenosis without a healthy control group (31), or a healthy data set only (4), stating the importance of performing such a comparative analysis between healthy subjects and CVD patients (4,18,31). Another study compares the vessel tortuosity in healthy subjects versus patients with intracranial artery atherosclerosis (9) but performed this manually and visually using 2D slices of time of flight (TOF) MRA lacking 3D segmentation or feature extraction, which could lead to miscalculation of the vascular features.

**Table 1.** A summary of related work in the literature on cerebral vascular segmentation. The geometric feature extraction column indicates whether the paper presented any geometric features of the vasculature and the skeletonization column indicates whether this method obtains the centerline and diameter information needed for CFD and mesh reconstruction. The last two columns specify the corresponding validation protocol and the major limitations which we tried to address in our method.

| Authors | Method | Modality | | (Centerline + Diameter information) | Geometric Feature Extraction | Validation Protocol | Major Limitations |
|---|---|---|---|---|---|---|---|
| Flasque et al.(32) | Centerline tracking and modeling | MRA | Manual or Semi-Automatic | ✓ | ✗ | Phantom | Manual intervention needed |
| Passat et al. (26) | Atlas registration with anatomical modeling and hit-or-miss transform | PC-MRA | | ✓ | ✗ | Manual | Manual intervention needed |
| Chen et al. (31) | Semi-automated Open-Curve Active Contour Vessel Tracing | 3D MRA | | ✓ | ✓ | Manual | Some manual intervention needed, only tested on patients with intracranial arterial stenosis |
| Gao et al. (33) | Statistical model analysis and curve evaluation | MRA | | ✗ | ✗ | Manual | Intensity based statistical analysis & local curve evaluation resulting in under-segmentation |
| Wright et al. (4) | Neuron_Morpho plugin ImageJ for segmentation, morphometric analysis & feature extraction | MRA | | ✓ | ✓ | NA | Insufficient Validation |
| Hsu et al. (18) | Multiscale composite filter & mesh generation | MRA | Fully Automatic | ✓ | Limited | Manual, phantom | Not tested on CT data, limited features |
| Wang et al. (34) | Otsu and Gumbel distribution based threshold | MRA | | ✗ | ✗ | Manual | Misclassification of skull pixels, under-segmentation of small vessels |
| Chen et al. (7) | Deep learning 3D U-Net architecture without manual annotation | MRA (CTA for training data) | | ✗ | ✗ | Manual | Thresholding based filtering to generate training data, insufficient validation |
| Meijs et al. (12) | Random forest classifier with local histogram features | 4D CT | | ✗ | ✗ | Manual | No geometrical information, manual validation |
| Zhao et al. (21) | Weighted Symmetry Filter | MRA, Retinal images | | ✗ | ✗ | Manual, phantom | No skeleton or geometrical information |

In this work, we propose an automated method for cerebral vascular segmentation that does not require initializing seed points or manual intervention and can be applied to any imaging modality; we test our method on both MRA and CTA data. Our method accounts for the differences in intensity inhomogeneities and tissue/bone contrast between MR and CT data and includes extensive validation using a realistic 3D phantom of the Circle of Willis. By skeletonizing the segmented vasculature, we extract global and regional geometric features of the vessel network to characterize the structure of the vascular tree. Finally, we present a quantitative comparison of the geometric features of the cerebral vascular tree between healthy subjects and stroke patients, to understand and quantify the structural differences in the vasculature caused by ischemic stroke, the most devastating cerebrovascular disease. The proposed method could be used to study any vascular malformations and diseases such as Alzheimer's, cerebral aneurysms, and stroke among others.

## Methods

In this section, we first detail the steps in our automatic segmentation and feature extraction algorithm: (1) pre-processing and vessel enhancement, (2) binarization and (3) skeletonization and feature extraction. We then present the validation of our method on a 3D phantom of the Circle of Willis (CoW) region of the vasculature, followed by the application of the proposed method to an MRA imaging dataset of healthy subjects and a CTA dataset of stroke patients, to compare vascular changes due to disease. A schematic overview of the process is presented below (Figure 1).

**Pre-processing:** We first performed skull-stripping using Hounsfield-unit thresholding and location-based segmentation for both MRA and CTA data to remove the bright skull regions that affect segmentation, especially in the CT data (35). This step prevents non-vascular skull pixels from being falsely enhanced in the vessel enhancement step due to a tubular structure or higher intensity. If the skull- stripping step is not performed, the segmentation would end up erroneously including non-vascular pixels. The MRA data is less susceptible to this limitation than CTA since it uses TOF (time of flight) imaging. Then we chose the region of interest (ROI) by selecting slices from the head region resulting in about 100 slices in the MR and 150 slices in the CT dataset (based on our datasets with 0.6mm axial resolution). This step improves computation speed significantly and reduces noisy structures that reside outside the cerebral vasculature region of interest.

**Segmentation:** The first step in the segmentation process is the vessel enhancement or 'vesselness' filtering, which is performed to suppress non-vascular structures and highlight the vessel pixels. We developed a custom multi-scale Frangi vesselness filter (36) to obtain a probability map of the pixels belonging to the vascular network (MATLAB, Mathworks, MA). After inputting raw MR or CT DICOM images (or other image formats), we apply a 2D Hessian based filter which enhances blood vessel contrast and eliminates other structures (36). The Hessian filter can be described as a second-order partial derivative of the image intensity map, aimed at tracking the path of least curvature and preserving tubular structures. The eigenvalues of the Hessian matrix depend on the directional voxel spacing and provide information about the shape of the object in the image. Accordingly, we assign a probability score to every pixel of being on the vessel with the center pixel having the highest probability and higher intensity, since vessels can be considered as 3D tubular structures at varying scales. The pixels with a higher probability are more likely to belong on a vessel and retain a much higher intensity than the background, enhancing the vessel contrast.

In the vessel-enhancement step, Gaussian smoothing is also performed along with the filtering to further reduce noise. The variance of the Gaussian kernel used for filtering is chosen based on the expected diameters of the vessels, since that maximally suppresses the noise around the blood vessels in the second order directional derivatives obtained with the Hessian filter. The multi-scale nature of the filter allows us to set local as well as global parameters, with the capability of detecting vessels as small in diameter as the image resolution (0.54 mm). Additional mathematical background for obtaining the Vesselness probability map and for the Gaussian filtering can be found in the supporting materials of this document.

In our implementation, we use the following parameters for the Vesselness and Gaussian filtering: *Scale* is defined as the standard deviation of the Gaussian kernel used for the analysis, which should be close to the expected radius of the vessel. *Minimum (Maximum) Scale* is the minimum (maximum) expected vessel radius in pixels, at which the relevant structure is expected to be found, and *Number of Scales* is an empirical parameter to set the range of radii detected. With this method, we can detect smaller vessels with the diameter in the range of the image resolution of 0.54mm. The Gaussian filtering for noise reduction is performed for different variances (σ), close to the expected vessel diameter calculated as:

$$Variance\ (\sigma) = \frac{(Maximum\ Scale - Minimum\ Scale)}{Number\ of\ Scales\ - 1}$$

The auxiliary scalar function value at every pixel is normalized to the maximum intensity of the image. And hence depends on depends on the gray-scale range of the input image. Half the value of the maximum Hessian norm has proven to work in most cases. This is chosen based on the input image dataset and was set to 500. For our study, we set Minimum Scale to 1 pixel, maximum scale to 20 pixels and the number of scales to 10, attributing to the varying vessel diameters in the cerebral vasculature.

Binarization: We segment the grayscale vascular probability map into a binary network of vessels using the 'Chan-Vese' Active Contours method (37). This method detects objects or regions of interest in an image based on parametric curve evolution and can iteratively detect objects without a gradient-defined boundary, using energy minimization. This segmentation technique, in combination with the pre-processing and multi-scale vesselness filtering provides a comprehensive 3D binary mask of the vasculature to be used for vectorization and feature extraction. The default number of iterations of active contours was set to 750 and a binary mask specifying rectangular ROI boundaries around the brain was provided as input. The algorithm then moves the mask to locate the object from the Vesselness images based on the specified number of iterations, without the need of any manual initialization or seed points. Due to the higher intensity of the vessel pixels and suppression of other structures in the preprocessing and vessel enhancement steps, the active contours automatically trace the vascular network, providing a binary volume.

For additional noise reduction, we multiply the vesselness map with a 3D mask before binarizing, to eliminate any stray skull/edge pixels, which also removed the venous structure(s) on the surface. Post binarization, to eliminate dangling structures or remnant noise from the segmentation, we perform an area opening operation (38) to discard disconnected segments smaller than a specified length using 3D 26-point connectivity. This gives us a 3D connected binary network of the cerebral arteries.

**Skeletonization and Reconstruction:** The binary vessel map was used to create a connected vascular network from which we extracted the geometric features corresponding to the entire vessel tree. We used medial axis thinning (39) to obtain the centerlines of the binary map and calculate the radius and angle(s) at each point on the centerline. Since we know the 3D centerline representation and the corresponding voxel indices on the centerline, we used the connected segments to calculate the 3D angles (axial and sagittal) between these points. To obtain the radius at each point on the centerline, we calculated the geodesic distance map (40) between center pixels and the boundary pixels and take the shortest distance between them as the radius at that point. To perform this, we first traced the exterior boundaries of the objects in the ROI as well as the inner edges of any 'holes' present using the Moore neighborhood tracing method (41,42) and then calculated the distance from the center pixel using the geodesic distance method. This provides us the radius of the vessel at every point on the centerline. Once we obtained the centerline network with precise radii at every point, we define a 'branching node' of the vascular tree as a point which is connected to three other points in 3D space, i.e., a bifurcation. After identifying all the branching nodes, we calculated the length of each vessel segment, defined as a series of connected points between two neighboring branching nodes. We then obtained quantifiable metrics of the vascular geometry in terms of these connected vessel segments. With this comprehensive information about the cerebral vascular tree, we reconstruct the arterial vasculature using the centerlines, radius and angular information by constructing 3D circles along the vessel centerline to form a 3D volume.

**Validation:** The accuracy of any vascular extraction method is determined by the precision of segmentation and the ability of the vessel enhancement and noise suppressing techniques. We extensively validated our method using a 3D vascular phantom of the Circle of Willis. This phantom establishes a physiologically realistic ground truth of the major arteries in the brain, against which the performance of our method was quantitatively evaluated since we know the geometrical properties of it. We 3D printed and CT scanned the physical 3D phantom to replicate human CTA data and ran our segmentation algorithm on the scan. Then we compared the reconstructed 3D geometry with the 'ground truth' (original binary phantom STL) and performed error analysis to quantify the performance of the segmentation algorithm. The validation metrics used were Pearson's correlation coefficient (43), Dice similarity coefficient (44), Modified Hausdorff Distance (45), and Surface distance image registration error (46).

Performance with CT image acquisition noise: Having validated our algorithm against the physical phantom, we ran the algorithm on the same phantom with varying levels of added Poisson and Gaussian noise, to mimic the added noise in a CT scan. This allowed us to predict how our algorithm would be affected by CT noise in a phantom study, allowing us to compare against ground truth measurement. To further evaluate and quantify how our method is affected by the presence of actual CT noise, we

3D printed the phantom and acquired a CT scan of the printed model at 0.6mm3. We then ran our algorithm on the CT data and compared the performance with the results from the original phantom data. We used a proprietary material by Stratasys Ltd. for 3D printing, known as Somos ® NeXt, which is a commonly used standard in literature for medical phantoms (3,47). Furthermore, we studied the dependence of the quality of segmentation and reconstruction on the image resolution as well. We down-sampled the phantom's image resolution from 0.48☐0.6☐0.6 mm$^3$ to 0.8☐0.8☐1.03 mm$^3$ and then to 1.12☐1.43☐1.5 mm$^3$ and then performed the segmentation and reconstruction algorithms to assess the ideal or minimum resolution needed for efficient and comprehensive reconstruction of the vascular network. We then compared this against the standard MR and CT resolution currently in practice.

Comparison with existing methods: For further validation, we quantitatively compared the results from our segmentation to the currently existing methods for vessel enhancement and extraction from popular image processing software such as ImageJ/FIJI (48).  Some of the existing algorithms which can be used to enhance vessels and create a binary map are auto/manual local thresholding, such as Renyi Entropy based thresholding (49) and Phansalkar Thresholding (50), Seeded Region Growing Segmentation (51), and Trainable Weka Segmentation (52). We applied these methods in the FIJI environment, along with ImageJ's implementation of Frangi Vesselness filtering.

**Geometric Feature Extraction:** With the comprehensive information contained in the skeletonized segments about the measurements of diameter, centerline points, angles, bifurcation points and branching structure, we then calculated the global and local morphometric features of the complete vascular tree. The features calculated were as follows: 1) *total length* of the vessel network calculated by summing the length of all the skeletal segments, 2) *total number of branches*, where a branch was defined as a sequence of points along the vessel starting at a bifurcation node and ending either at the next bifurcation or at the last point on the vessel (in the case of a terminating branch), 3) *average and maximum branch length*, defined as the mean and maximum geodesic length of all branches in the network, 4) *average diameter* of all points on the centerline, 5) *total vessel volume*, calculated by considering the vessels as cylinders with a varying diameter along the total length, 6) *fractal dimension,* determined using the box counting method based on the Minkowski-Bouligand dimension (53–55), which provides a measure of morphological complexity in the cerebral vasculature, and 7) *vessel tortuosity*, defined using the sum of angles measurement (SOAM) (56), and calculated as the sum of all the angles between sets of 3 points on the centerline divided by the total length (SOAM). This feature has been linked to potential vascular pathology such as atherosclerosis and even brain tumor malignancy and can be used to study the changes in vessel structure in such cerebrovascular diseases (9,57).

**Healthy vs. Stroke Comparison:** The geometric properties described in the previous section were obtained for the cerebral trees of two groups of data: (i) MRA scans of healthy subjects (n=10, age = 30±9), and (ii) CTA scans of stroke patients (n=10, age = 68±11). Both groups include both male and female subjects in an equal ratio. For the healthy dataset, subjects with any history of hypertension, diabetes or any head trauma were excluded. The stroke patient dataset consists of CTA images of older adults within the same age range and with major vessel occlusion in the M1, M2 or ICA sites, which are the most common sites of vessel occlusion in an ischemic stroke.

Medical imaging protocols: A set of Computed Tomography Angiography (CTA) scans from stroke patients, acquired at 0.43☐0.43☐0.62 mm$^3$ resolution using the GE Lightspeed scanner at 120KV was provided by the Stanford University Medical Center. All the CTA data was acquired after bolus injection of 90-120ml contrast media with the injection rate of 4-5ml/s (Omnipaque, 350mg/ml) at the Centre Hospitalier Universitaire Vaudois, Lausanne, Switzerland. From this dataset, scans of 10 stroke patients were used for the comparative study with healthy subjects.

The MIDAS Magnetic Resonance Angiography (MRA) database was used for the healthy subject data. The TOF MR brain images from 109 healthy volunteers acquired at 0.5☐0.5☐0.8 mm$^3$ were collected and made available by the CASILab at The University of North Carolina at Chapel Hill and were distributed by the MIDAS Data Server at Kitware, Inc. (8). For this study, we utilized data from 10 subjects.

We compared each extracted feature between the two groups using a paired t-test and ANOVA. We expected to find the average diameter, volume and total length in stroke patients to be smaller due to a major vessel being occluded with a higher tortuosity and fractality since these have been shown to be indicators of vascular pathology (6,9,58). We also hypothesized that the number of branches would be lesser in the stroke data due to vessel occlusion but there is also contradictory information in literature regarding this as the vasculature tends to sprout additional smaller branches to compensate for the stroke (11,59). The results from the quantitative comparison can be found in table 4. Results are presented as the mean ± standard deviation and by group.

## Results

**Validation:** Validation studies using the 3D phantom show that our algorithm detects vessels accurately with a Dice similarity coefficient (DSC) of 84%, Pearson's correlation coefficient (PCC) of 83.4% and a modified Hausdorff distance of at most 3 pixels (Figure 1 and Table 2). We can see that the error is confined to the surface pixels of the vessels due to slight thinning but overall, the algorithm detects every vessel segment with a high accuracy and allows for a comprehensive reconstruction with the loss being restricted to 1-3 pixels confined to the surface.

In our validation scheme, we also demonstrate the impact of noise on our algorithm. The noise added phantom with varying levels (10%, 20%) of Poisson and Gaussian noise was also reconstructed with 83% DSC along with the CT scanned 3D printed phantom with inherent CT acquisition noise (Table 2), providing comparison against ground truth for data with CT noise. This establishes the accuracy of our method when used with existing noise induced by image acquisition and reconstruction.

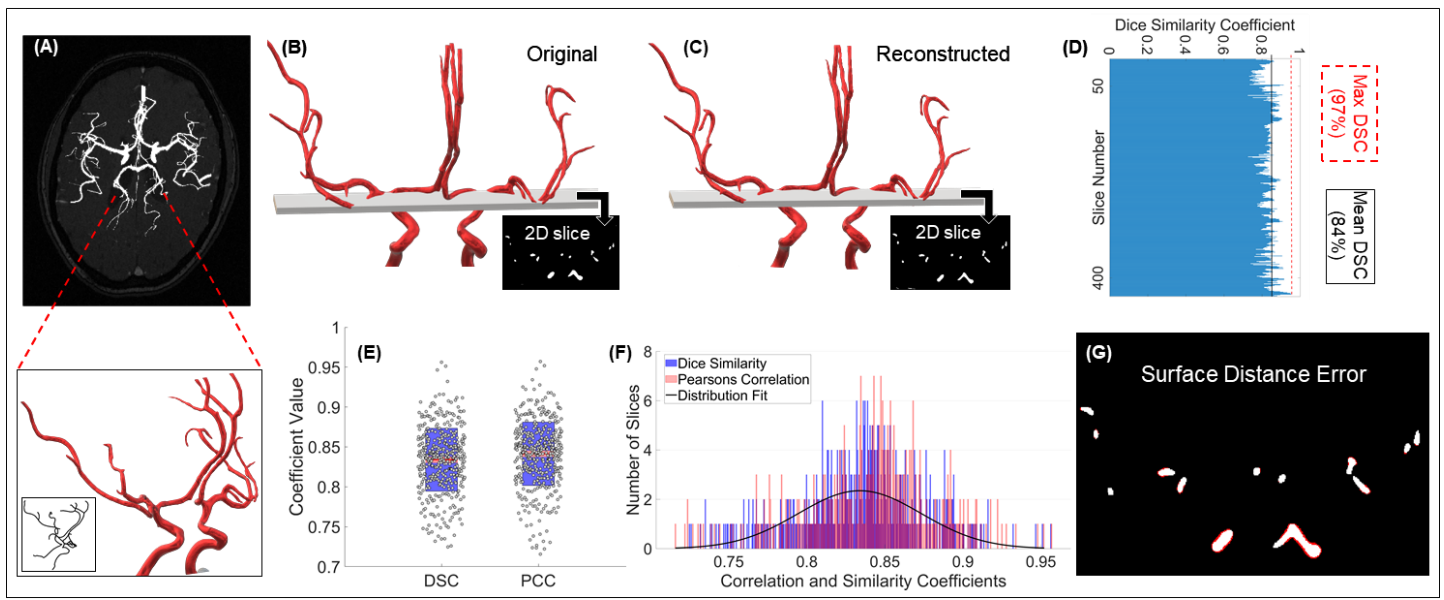

**Figure 1.** Validation of Segmentation using a 3D phantom. (A) Extracted vasculature overlaid on TOF image of an axial slice of the brain and the 3D CoW phantom shown relative to the cerebral vasculature along with its extracted centerlines (embedded) for a visual representation (B) Original 3D phantom of the CoW with a 2D slice showing a cross section, (C) The 3D volume reconstructed using the CT scan of the 3D printed phantom with the corresponding 2D cross section shown for visual comparison of the 2D slices, (D) DSC per 2D slice corresponding to the reconstruction of the 3D phantom(s) on the left post CT scanning the 3D printed phantom, (E) Box plots of the DSC and the PCC showing the data points laid over a 95% confidence interval, along with (F) Histogram of the DSC and PCC demonstrating the accuracy of the segmentation along with the distribution fit for the DSC, lastly, (G) Corresponding 2D slice (shown in B&C) showing an overlap of the original and segmented cross section, indicating the error for visualization (the 'error pixels' can be seen in red).

**Table 2.** Validation and error analysis results for the 3D phantom as well as the quantification of performance with varying levels of added noise. Lastly, segmentation result using CT scan images of the 3D printed phantom to account for CT induced noise and comparison with ground truth data.

|  | Dice Similarity Coefficient (%) | Pearson's correlation (%) | Modified Hausdorff Distance (pixels) |
| --- | --- | --- | --- |
| Phantom | 84.3 | 83.9 | 3 |
| Phantom + 10% noise | 84.7 | 84.2 | 3 |
| Phantom + 20% noise | 83.7 | 83.1 | 3 |
| 3D print + CT of phantom | 84.6 | 84.5 | 2 |

We see a slight increase in DSC and PCC with added noise (Table 2) since Gaussian and Poisson noise add grayscale pixels randomly around the ROI, falsely appending a few pixels 'missed' in the segmented volume as the error is confined only to the pixels on the surface.

We also studied the effects of image resolution in the vessel extraction and reconstruction, inferring that at worse resolutions below the standard CTA and MRA resolution (~0.5 to 0.6 mm), there are discontinuities in the segmented binary map and an over/under estimation of the radius at certain points. A graph showing the fall in the Dice Similarity Coefficient with declining resolution below clinical standard is presented in supplemental material of this text. The quantitative comparison of results from ImageJ/FIJI using currently existing segmentation protocols against our method using the 3D phantom as the ground truth showed that our method of binarization combined with our implementation of the Frangi vesselness filtering outperforms the FIJI algorithms. This can be seen in Figure 2. Furthermore, Table 3 reports the comparison of the methods against the ground truth in terms of Dice Similarity Coefficient, Pearson's Correlation Coefficient and the Modified Hausdorff Distance.

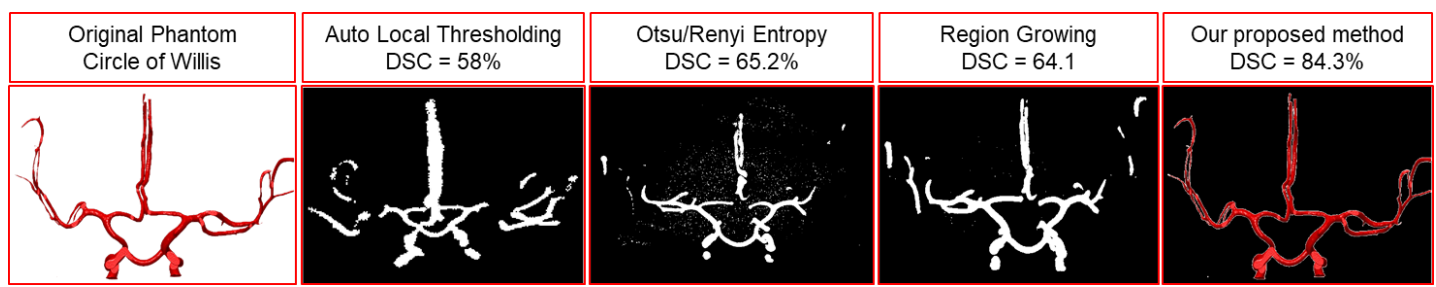

**Figure 2.** The segmentation results from other existing methods (Implement in FIJI/ImageJ) along with the current proposed method for visual comparison.

**Table 3.** Comparison with other methods. Segmentation results and subsequent error analysis using other existing methods – performed using the commonly used ImageJ/FIJI software for comparison against results from our algorithm, reported in Table 2.

| Segmentation method | Dice Similarity Coefficient (%) | Pearson's correlation (%) | Modified Hausdorff Distance (pixels) |
|---|---|---|---|
| Auto local thresholding | 58 | 57.6 | 5 |
| Region growing | 64.1 | 63.3 | 4 |
| Otsu/Renyi Entropy | 65.2 | 66.4 | 4 |
| Proposed method | 84.3 | 83.9 | 3 |

**Segmentation:** The result from each step in the entire process is shown in Figure 3. We can see the vesselness map obtained by the vessel enhancement filtering after pre-processing and the subsequent anisotropic diffusion. It can be seen clearly that the MCA (Middle Cerebral Artery) and the Internal Carotid Arteries (ICA) have the highest probability value(s) in the vesselness map and hence appear the brightest. The smaller vessels such as the communicating arteries can be seen but are faint in comparison. Overall, we can very clearly see the enhanced contrast in the vessels and the suppression of other structures. The binarized map obtained using the active contours segmentation shows how the automatic algorithm works to pick up the faint segments for better connectivity which thresholding-based algorithms tend to miss, without the need for manually placing seed points. We can see how the 3D volume is reconstructed using the extracted centerline, diameter and angular information in the skeletonized visualization of the centerline trajectory and corresponding diameters in Figure 3 (D).

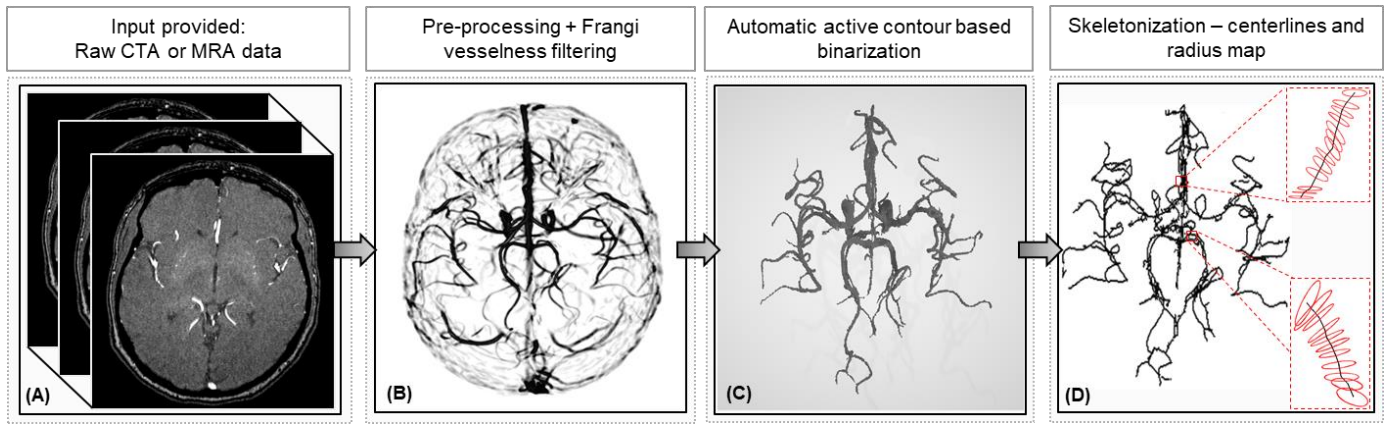

**Figure 3.** Vessel segmentation and skeletonization: (A) raw stack of 2D MR/CT images, (B) vesselness map obtained after pre-processing and filtering, (C) binary volume obtained using active contours segmentation, and (D) skeleton of the cerebral vasculature centerlines and surface cross-sections depicted by 3D circles and corresponding diameter values from segmentation.

The segmented vasculature of the stroke subject clearly shows the MCA M1, M2 segments being cut off at the point marked by the arrow in Figure 4 in which the healthy and stroke vasculature can be visually compared. Table 4 contains the quantitative information about the extracted geometrical features and a comparison between healthy vs. stroke vascular geometry.

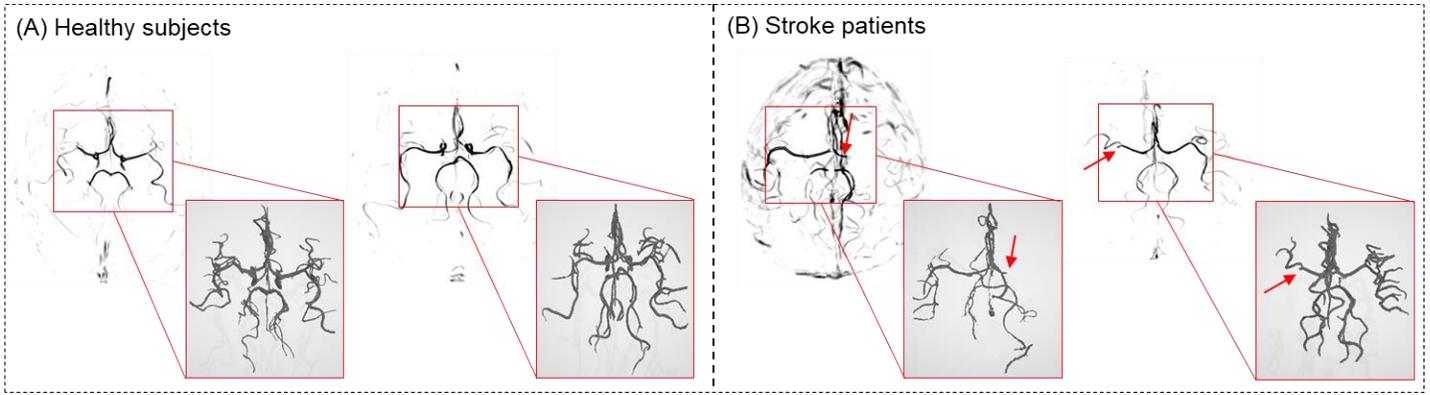

**Figure 4.** A visual comparison of the vesselness map and corresponding binary volume of the cerebral vasculature: (A) two healthy subjects, and (B) two stroke patients. The red arrows on the stroke image data depicts the location of the blood flow being cut off due to the stroke.

**Geometric Feature Extraction:** The results from the geometric feature extraction can be seen in Table 4. The Total Length', 'Total number of Branches', 'Average Diameter', 'Average and maximum Branch Length', 'Total Volume', 'Fractal Dimension' and 'Vessel Tortuosity' have been reported as the mean ± SD for the healthy subjects as well as the stroke patients. The data shows that the obtained values for the average diameter, length and branching of cerebral arterial trees agree with the values reported in literature (18,31).

**Table 4.** A comparison of geometric features of the cerebral vascular tree of healthy subjects vs. stroke patients. Values are presented as average ± standard deviation of 10 subjects in each group with the **bold font** highlighting features that are significantly different between groups (**p<0.05**).

|  | Averaged values from healthy subjects | Averaged values from stroke patients | *p* - value |
| --- | --- | --- | --- |
| Total length (m) | 2.88 ± 0.38 | 2.20 ± 0.67 | **0.005** |
| Number of branches | 125 ± 76 | 211 ± 75.69 | 0.051 |
| Average branch length (mm) | 14.81 ± 1.97 | 9.89 ± 2.07 | **<0.001** |
| Maximum branch length (mm) | 59.25 ± 10.78 | 59.38 ± 6.10 | 0.738 |
| Average diameter (mm) | 2.75 ± 0.37 | 2.18 ± 0.38 | **0.007** |
| Total volume (ml) | 40.18 ± 25.55 | 34.43 ± 21.83 | **0.013** |
| Fractal dimension | 1.36 ± 0.28 | 1.69 ± 0.20 | **0.007** |
| Tortuosity (rad/cm) | 3.24 ± 0.88 | 5.80 ± 0.92 | **<0.001** |

It can be seen from Table 4 that most of the extracted geometrical features are significantly different in stroke patients as compared to healthy subjects. As hypothesized, the total length, volume and average diameter of cerebral vessels in stroke patients was lower whereas the Fractal dimension, tortuosity was significantly higher. The number of branches and branch length had too high a variance to be judged as significantly different.

Discussion:

In this paper, we present a method to automatically segment and reconstruct cerebral vasculature without the use of seed points or initialization, obtaining a comprehensive binary network of the vessels. This method attempts to overcome the limitations of semi-automatic methods in literature which need manual intervention, with minimal pre-processing and no further post processing required. This method was tested on MRA as well as CTA data and can be used with any modality of imaging blood vessel contrast without being restricted to specific imaging protocols and can detect small vessels at the size of the image resolution (0.5mm) without any manual initialization or intervention needed. We validated our method through extensive and rigorous error analysis studies using a 3D phantom of the Circle of Willis. We also studied the effect of CT noise level in our algorithm. The phantom utilized for validation served as a ground truth and an ideal reference which allowed for quantification of the algorithm's performance, using Pearson's Correlation Coefficient, Dice Similarity, Modified Hausdorff Distance and Surface Distance Error. We compared our results against other existing methods and showed superior performance visually as well as with quantifiable metrics in Figure 2 and Table 3. This demonstrates that other methods such as auto local thresholding, entropy or class-based segmentation and other methods such as region growing, which need seed points don't achieve the segmentation accuracy of our method which combines an in-house implementation of Frangi filtering with active contours-based ROI tracing. In figure 4, it can be seen that in cases where the 'vesselness' map post Frangi filtering appears visually too faint (lower probability of those pixels than others due to thinner/low contrast vessels), the active contours can still trace those vessel pixels and they get picked up and included in the binary volume as the probability is non-zero and still higher than the background. Hence, this combination algorithm overcomes the challenges of inconsistent intensity values over the length scales of various vessels and outperforms other methods.

We further developed the algorithm to include automatic feature extraction of the vessels to characterize patient-specific cerebral vasculature. This algorithm skeletonizes the vascular network and extracts regional geometric features such as length, diameter, branching pattern, fractal dimension and tortuosity, which can be used to study the mechanism of vascular pathology and biometrics of structural changes in the cerebral vessels. Such an extensive analysis of vascular features has been very scarcely reported in literature with no other work presenting a comparison of healthy with pathological vasculature. Using this algorithm, we performed a comparative study of the vascular geometry in stroke patients vs. healthy subjects to quantify the structural changes in the cerebral vasculature induced by ischemic stroke, which is the largest contributor to death and disability due to cerebrovascular disease. The results show that the vascular geometry differs significantly between the two groups and to the best of our knowledge, this is the first study to present this quantitative comparison data using automated segmentation and skeletonization of vasculature. The features demonstrating significant differences between groups were total length (2.88 ± 0.38 m for healthy and 2.20 ± 0.67 m for stroke), volume (40.18 ± 25.55 ml for healthy and 34.43 ± 21.83 ml for stroke), tortuosity (3.24 ± 0.88 rad/cm for healthy and 5.80 ± 0.92 rad/cm for stroke) and fractality (box dimension1.36 ± 0.28 for healthy vs. 1.69 ± 0.20 for stroke) with ANOVA using $p < 0.05$. As we hypothesized prior to conducting this study, the stroke vasculature was found to have a lower volume and total length as well as smaller average diameter, however, the vascular network of stroke patients possessed higher tortuosity, fractality and branching. This is consistent with findings from literature regarding vessels being more complex and tortuous in stroke patients along with additional smaller branches forming for collateral flow. We conclude that the volume and total length are still lower because the newer collateral branches formed are smaller than the major vessel network missing in the segmented volume due to the occlusion.

It is important to note that aging has a significant impact on vascular impairment since the two groups presented here have a large different in their age range (30±9 years for healthy subjects and 68±11 years for stroke patients). Multiple studies have noted that aging results in an increase in arterial stiffness, arteriolar tortuosity and endothelial molecular dysfunction, potentially leading to hypo-perfusion (60). These alterations in the vasculature, in turn lead to pathophysiological manifestations such as atherosclerotic vascular diseases (i.e. stroke), aneurysms, vascular inflammation, hypertension and hemorrhages (61,62). Hence these structural changes in the vasculature found in stroke patients, are also directly correlated with the aging process.

This method provides a basis for a quantitative tool to study vascular pathology in various underlying cerebrovascular diseases as well as to accurately segment vasculature for visualization and assessment in the efficient diagnosis and treatment of stroke. Apart from these diagnostic and prognostic applications, reconstruction of patient-specific cerebrovascular network is a vital step for computational fluid dynamics (CFD) modeling studies which can analyze cerebral hemodynamics (13,63) and provide outcomes for various forms of vascular interventions. Important components of CFD studies include 3D reconstruction of the patient's vascular network and a knowledge of geometric features such as the centerlines, diameters and bifurcations (18,63).

**Limitations:** There could be discontinuities in the binary network if the original image dataset itself has inconsistent and poor contrast in some slices where the vessel cross section cannot be seen entirely. In such cases, we eliminated the corresponding vascular regions which were disconnected from the entire structure, in an attempt to only preserve the completely connected network. With a more comprehensive understanding of human cerebral vasculature, we would be able to implement a method to detect and correct for such discontinuities, especially for geometric feature analysis and CFD studies. Lastly, with this method and all other methods in literature, due to the limitations of imaging resolution, we still can't detect and segment microvasculature which would provide a greater insight into cerebral hemodynamics.

**Acknowledgements:** The study was supported by the National Institutes of Health (NIH) National Institute of Neurological Disorders and Stroke (NINDS) grant number R03NS108167 and Tech Launch Arizona UA18-140.